\documentclass[12pt,a4paper,landscape]{article}

\newcommand{\rd}{\mathrm{d}}  
\newcommand{\re}{\mathrm{e}}  

\newcommand{\R}{\mathbf{R}}   
\newcommand{\C}{\mathbf{C}}   
\newcommand{\Z}{\mathbf{Z}}   
\newcommand{\BS}{\mathbf{S}}   
\newcommand{\Rplus}{\mathbf{R}_{\ge 0}}

\newcommand{\bra}{\langle}
\newcommand{\ket}{\rangle}

\newcommand{\Lie}{\mathbf{g}}
\newcommand{\LieSO}{\mbox{\bf so}}
\newcommand{\LieU}{\mbox{\bf u}}

\def\dfrac#1#2{\displaystyle\frac{#1}{#2}}

\newcommand{\isoarrow}
{\mathop{\hbox to 5mm{\rightarrowfill}}
\limits^{\scriptstyle \sim}}

\newcommand{\mapright}[2]
{\mathop{\hbox to 1cm{\rightarrowfill}}
\limits^{\scriptstyle #1}_{\scriptstyle #2}}

\newcommand{\mapdown}[2]
{\Big \downarrow 
\llap {$\vcenter {\hbox{$\scriptstyle #1 \,$}}$ }
\rlap {$\vcenter {\hbox{$\scriptstyle #2   $}}$ }
}

\makeatletter
\@addtoreset{equation}{section}
\makeatother

\begin{document}
%
\baselineskip 6mm
\hfill
hep-th/0110015
\vspace{4mm}
\begin{center}
{\bf \Large 
Path Integrals on Riemannian Manifolds with Symmetry
and Stratified Gauge Structure}
\footnote{
Proceedings of 
The Third International Conference on Geometry, Integrability and Quantization,
which was held in June 14--23, 2001
at Sts. Constantine and Elena resort in Bulgaria.}
\vspace{4mm} \\
Shogo Tanimura
\\
{\it Department of Engineering Physics and Mechanics,
Kyoto University, \\
Kyoto 606-8501, Japan} \\
e-mail: tanimura@kues.kyoto-u.ac.jp
\vspace{8mm}\\
Abstract \\
\begin{minipage}[t]{140mm}
We study a quantum system in a Riemannian manifold $ M $ on which
a Lie group $ G $ acts isometrically.
The path integral on $ M $ is decomposed into a family of
path integrals on a quotient space $ Q=M/G $
and the reduced path integrals are completely classified 
by irreducible unitary representations of $ G $.
It is not necessary to assume that
the action of $ G $ on $ M $ is either free or transitive.
Hence the quotient space $ M/G $ may have orbifold singularities.
Stratification geometry,
which is a generalization of the concept of principal fiber bundle,
is necessarily introduced
to describe the path integral on $ M/G $.
Using it we show that
the reduced path integral is expressed as a product of three factors;
the rotational energy amplitude, 
the vibrational energy amplitude,
and the holonomy factor.
\end{minipage}
\end{center}
\vspace{4mm}
\section{Basic observations and the questions}
Let us consider the usual quantum mechanics 
of a free particle in the one-dimensional space $ \R $.
A solution for the initial-value problem
of the Schr{\"o}dinger equation
\begin{equation}
	i \frac{\partial}{\partial t} \phi(x,t) 
	= - \frac{1}{2} \frac{\partial^2}{\partial x^2} \phi(x,t)
	= \frac{1}{2} \Delta \phi(x,t)
	\label{Schr}
\end{equation}
is given by
\begin{equation}
	\phi(x,t) = \int_{-\infty}^{\infty} \rd y \, K(x,y;t) \phi(y,0) 
\end{equation}
with the propagator
\begin{equation}
	K(x,y;t) 
	= \bra x | \re^{-\frac{i}{2} t \Delta } | y \ket
	= \frac{1}{ \sqrt{2 \pi i t} } 
	\exp \left[ \frac{i}{2t} (x-y)^2 \right].
\end{equation}
Their physical meanings are clear;
the wave function $ \phi(x,t) $ represents probability amplitude
to find the particle at the location $ x $ at the time $ t $.
The propagator $ K(x,y;t) $ represents transition probability amplitude
of the particle to move from $ y $ to $ x $ in the time interval $ t $.

If the particle is confined in the half line $ \Rplus = \{ x \ge 0 \} $,
we need to impose a boundary condition on the wave function $ \phi(x,t) $
at $ x=0 $ to make the initial-value problem (\ref{Schr})
have a unique solution.
As one of possibilities 
we may chose the Neumann boundary condition
\begin{equation}
	\frac{\partial \phi}{\partial x} (0,t) = 0.
\end{equation}
Then the solution of (\ref{Schr}) is given by
\begin{equation}
	\phi(x,t) = \int_{-\infty}^{\infty} \rd y \, K_N (x,y;t) \phi(y,0) 
\end{equation}
with the corresponding propagator 
\begin{equation}
	K_N (x,y;t) = K(x,y;t) + K(-x,y;t).
\end{equation}
Physical meaning of the propagator $ K_N(x,y;t) $ is obvious;
the first term $ K(x,y;t) $ represents propagation of a wave 
from $ y $ to $ x $ while
the second term $ K(-x,y;t) $ represents propagation of a wave 
from $ y $ to $ -x $, which is the mirror image of $ x $.
Thus the Neumann propagator $ K_N(x,y;t) $ is a superposition
of the direct wave with the reflected wave.

As an alternative choice
we may impose the Dirichlet boundary condition
\begin{equation}
	\phi (0,t) = 0.
\end{equation}
Then the solution of (\ref{Schr}) is given by
\begin{equation}
	\phi(x,t) = \int_{-\infty}^{\infty} \rd y \, K_D (x,y;t) \phi(y,0) 
\end{equation}
with the corresponding propagator 
\begin{equation}
	K_D (x,y;t) = K(x,y;t) - K(-x,y;t).
\end{equation}
Thus 
the Dirichlet propagator $ K_D(x,y;t) $ is also a superposition
of the direct wave with the reflected wave
but reflection changes the sign of the wave.

The half line $ \Rplus $ can be regarded as an orbifold $ \R / \Z_2 $.
In the above discussion
we assumed existence of the propagator $ K(x,y;t) $ in $ \R $
and constructed the propagators in $ \R / \Z_2 $ from $ K(x,y;t) $.
There are two inequivalent propagators;
the Neumann propagator $ K_N(x,y;t) $ 
obeys the trivial representation of $ \Z_2 $
whereas
the Dirichlet propagator $ K_D(x,y;t) $ 
obeys the defining representation of $ \Z_2 = \{ +1, -1 \} $.

Now a question arises;
how is a propagator in a general orbifold $ M/G $ constructed?
Here $ M $ is a Riemannian manifold
and $ G $ is a compact Lie group that acts on $ M $ by isometry.
Such an example is easily found;
we may take $ M = \BS^2 $ and $ G = U(1) $.
Then the quotient space is $ M/G = [-1,1] $,
which has two boundary points.

Let us turn to another aspect of the propagator, namely, 
the path-integral expression of the propagator.
{}For the general Schr{\"o}dinger equation
\begin{equation}
	i \frac{\partial}{\partial t} \phi(x,t) 
	= H \phi(x,t)
	= - \frac{1}{2} \frac{\partial^2}{\partial x^2} \phi(x,t) 
	+ V(x) \phi(x,t),
	\qquad x \in \R,
\end{equation}
its solution is formally given by
\begin{equation}
	\phi(x,t) = \int_{-\infty}^{\infty} \rd y \, K(x,y;t) \phi(y,0).
\end{equation}
The propagator satisfies the composition property
\begin{equation}
	K(x'',x;t+t') = 
	\int_{-\infty}^{\infty}  \rd x' \, K(x'',x';t') K(x',x;t).
\end{equation}
By dividing the time interval $ [0,t] $ into short intervals we get
\begin{equation}
	K(x_N, x_0; t) 
	=
	\int_{-\infty}^{\infty} \! \cdots
	\int_{-\infty}^{\infty} \rd x_{N-1} \cdots \rd x_1
	K(x_N, x_{N-1}; \epsilon) \cdots
	K(x_{1}, x_{0}; \epsilon) 
	\quad
\end{equation}
with $ t = N \epsilon $. 
For a short distance and a short time-interval
the propagator asymptotically behaves as
\begin{equation}
	K( x + \Delta x, x; \Delta t) 
	\sim 
	\frac{1}{ \sqrt{2 \pi i \Delta t} } 
	\exp 
	\left[ 
		\frac{i}{2} 
		\left( \frac{\Delta x}{\Delta t} \right)^2 \!\! \Delta t
		- i V(x) \Delta t
	\right].
\end{equation}
Then ``the limit $ N \to \infty $'' gives an infinite-multiplied integration,
which is called the path integral,
\begin{equation}
	K(x', x; t) 
	= 
	\int_x^{x'} \! {\cal D}x \, \re^{ i \int L \, \rd s }
	=
	\int_x^{x'} \! {\cal D}x 
	\exp
	\left[
		i \! \int_0^t \! \rd s 
		\left( \frac{1}{2} \dot{x}(s)^2 - V(x(s)) \right)
	\right].
\end{equation}
In a rigolous sense,
the limit $ N \to \infty $ dose not exists but
physicists use this expression for convenience.
Philosophy of the path integral is symbolically written as
\begin{equation}
	\mbox{propagation of the wave \quad} 
	=
	\sum_{\mbox{ trajectories }} \mbox{motion of the particle}.
\end{equation}
We can construct the path integral on the half line $ \Rplus = \R/\Z_2$
as well:
\begin{eqnarray}
&&	K_N (x', x; t) 
	=
	\sum_{n=0}^\infty
	\int_x^{x'} \! {\cal D}x \, \re^{ i \int L \, \rd s },
	\\
&&	K_D (x', x; t) 
	=
	\sum_{n=0}^\infty (-1)^n
	\int_x^{x'} \! {\cal D}x \, \re^{ i \int L \, \rd s },
\end{eqnarray}
where the summations are taken with respect to the number of reflections
of the trajectory at the boundary $ x=0 $.

Now another question arises;
what is the definition of path integrals on a general orbifold $ M/G $?
Our main concerns are propagators and path integrals in $ M/G $.

\section{Reduction of quantum system}
When a quantum system has symmetry,
it is decomposed into a family of quantum systems that are defined
in the subspaces of the original.
Here we review the reduction method \cite{TI} of quantum system.

A quantum system $ ( {\cal H}, H ) $ is defined by a pair of 
a Hilbert space $ {\cal H} $ and
a Hamiltonian $ H $, which is a self-adjoint operator on $ {\cal H} $.
Symmetry of the quantum system is specified by
$ ( G, T ) $,
where 
$ G $ is a compact Lie group
and 
$ T $ is a unitary representation of $ G $ over $ {\cal H} $.
Symmetry implies that $ T(g) H = H T(g) $ for all $ g \in G $.
The compact group $ G $ is equipped with 
the normalized invariant measure $ \rd g $.

To decompose $ ( {\cal H}, H ) $ into a family of reduced quantum systems,
we introduce
$ ( {\cal H}^\chi, \rho^\chi ) $,
where
$ {\cal H}^\chi $ is a finite dimensional Hilbert space 
of the dimensions $ d^\chi = \dim {\cal H}^\chi $.
Besides, $ \rho^\chi $ is
an irreducible unitary representation of $ G $ over $ {\cal H}^\chi $.
The set $ \{ \chi \} $ labels all the inequivalent representations.
{}For each $ g \in G $,
$ \rho^\chi(g) \otimes T(g) $ acts on $ {\cal H}^\chi \otimes {\cal H} $
and defines the tensor product representation.
The {\it reduced Hilbert space} is defined as
the subspace of the invariant vectors of $ {\cal H}^\chi \otimes {\cal H} $,
\begin{equation}
	( {\cal H}^\chi \otimes {\cal H} )^G
	:= 
	\{ \psi \in {\cal H}^\chi \otimes {\cal H} 
	\, ; \,
	\forall h \in G, \,
	( \rho^{\chi}(h) \otimes T(h) ) \psi = \psi
	\}.
	\label{invariant space}
\end{equation}
A set $ \{ e^\chi_1, \dots, e^\chi_d \} $ is an orthonormal basis of $ {\cal H}^\chi $.
Then the {\it reduction operator}
$ S^\chi_i : {\cal H} \to ({\cal H}^\chi \otimes {\cal H})^G $
is defined by
\begin{equation}
	f \in {\cal H} \mapsto
	S^\chi_i f
	:= 
	\sqrt{d^\chi} \int_G \rd g \,
	( \rho^\chi(g) e^\chi_i ) \otimes ( T(g) f ).
\end{equation}
\\
{\bf Theorem 2.1.}
{\it
$ S^\chi_i $ is a partial isometry. Namely,
$ (S^\chi_i)^* S^\chi_i $ is an orthogonal projection operator 
acting on $ {\cal H} $ 
while
$ S^\chi_i (S^\chi_i)^* $ is the identity operator on
$ ({\cal H}^\chi \otimes {\cal H})^G $.
}
\vspace{3mm}
\\
{\bf Theorem 2.2.}
{\it
The family of the projections
$ \{ (S^\chi_i)^* S^\chi_i \} $ 
forms a resolution of the identity as
\begin{equation}
	\sum_{\chi, i} (S^\chi_i)^* S^\chi_i = I_{\cal H}.
\end{equation}
Hence, the Hilbert space is decomposed as
\begin{equation}
	{\cal H} 
	=
	\bigoplus_{\chi, i} \, \mbox{\rm Im} \, (S^\chi_i)^*  S^\chi_i
	\cong
	\bigoplus_{\chi, i} \, ({\cal H}^\chi \otimes {\cal H})^G
\end{equation}
and this decomposition is compatible with action of the Hamiltonian.
Namely, we have the commutative diagram
\begin{equation}
	\begin{array}{ccc}
	  {\cal H}
	& \mapright{ S^\chi_i }{}
	& ( {\cal H}^\chi \otimes {\cal H})^G 
	\\
	  \mapdown{ H \,}{}
	& 
	& \mapdown{}{ %
	  \, \mbox{\scriptsize \rm id} \otimes H }
	\\
	  {\cal H}
	& \mapright{ S^\chi_i }{}
	& ( {\cal H}^\chi \otimes {\cal H})^G 
	\end{array}
\end{equation}
Then $ ( ( {\cal H}^\chi \otimes {\cal H})^G, \mbox{\rm id} \otimes H ) $ 
defines a reduced quantum system.
}
\vspace{3mm}

The projection
$ P^\chi : 
{\cal H}^\chi \otimes {\cal H} \to ({\cal H}^\chi \otimes {\cal H})^G $
onto the reduced space is defined by
\begin{equation}
	P^\chi 
	:= \int_G \rd g \, \rho^\chi(g) \otimes T(g).
\end{equation}
The {\it reduced time-evolution operator} of the reduced system is
\begin{equation}
	U^\chi := P^\chi ( \mbox{\rm id} \otimes \re^{-iHt} ).
	\label{red time}
\end{equation}
Theorems 2.1 and 2.2 are easily proved 
by application of the Peter-Weyl theorem,
which states that
the set 
of the matrix elements of irreducible unitary representations
$ \{ \sqrt{d^\chi} \, \rho^\chi_{ij} (g) \}_{\chi,i,j} $ 
forms a complete orthonormal set of $ L_2(G) $.
Our main purpose is to give a path-integral expression
to the time-evolution operator $ U^\chi $.
To describe it we need to prepare some related notions.

Assume that the base space $ M $ is equipped with the measure $ \rd x $.
Then 
the space of the square-integrable functions $ L_2(M) $ 
becomes a Hilbert space $ {\cal H} $.
Moreover, assume that the compact Lie group $ G $ acts on $ M $ 
with preserving the measure $ \rd x $.
Then
$ g \in G $ is represented by the unitary operator 
$ T(g) $ on $ f \in L_2(M) $ by
\begin{equation}
	( T(g) f ) (x) := f(g^{-1}x).
	\label{T}
\end{equation}
Let $ p : M \to Q = M/G $ be the canonical projection map.
Then 
a measure $ \rd q $ of $ Q = M/G $ is induced by the following way.
Let $ \phi(q) $ be a function on $ Q $ such that
$ \phi(p(x)) $ is a measurable function on $ M $.
The induced measure $ \rd q $ of $ Q $ is then defined by
\begin{equation}
	\int_Q \rd q \, \phi(q) := \int_M \rd x \, \phi(p(x)).
\end{equation}
On the other hand,
suppose that the time-evolution operator $ U(t) := \re^{-iHt} $ 
is expressed in terms of an integral kernel 
$ K: M \times M \times \R_{>0} \to \C $ as
\begin{equation}
	( U(t) f ) (x) = \int_M \rd y \, K(x,y;t) f(y)
	\label{U(t)}
\end{equation}
for any $ f(x) \in L_2(M) $.

Let us turn to the reduced Hilbert space (\ref{invariant space})
and characterize it for the case $ {\cal H} = L_2(M) $.
A vector $ \psi \in {\cal H}^\chi \otimes L_2(M) $ 
can be identified with a measurable map
$ \psi : M \to {\cal H}^\chi $.
The tensor product $ \rho^\chi(g) \otimes T(g) $ acts on $ \psi $ as
\begin{equation}
	( (\rho^\chi(g) \otimes T(g)) \psi ) (x)
	=
	\rho^\chi(g) \psi (g^{-1} x),
	\quad
	g \in G
\end{equation}
via the definition (\ref{T}).
The definition (\ref{invariant space}) of the invariant vector 
$ \psi \in ({\cal H}^\chi \otimes L_2(M))^G $ implies
\begin{equation}
	( (\rho^\chi(g) \otimes T(g)) \psi ) (x)
	=
	\rho^\chi(g) \psi (g^{-1} x)
	=
	\psi(x),
\end{equation}
which is equivalent to
\begin{equation}
	\psi(gx) = \rho^\chi(g) \psi (x).
\end{equation}
A function $ \psi : M \to {\cal H}^\chi $
satisfying the above property is called an {\it equivariant function}.
Hence the reduced Hilbert space is identified with the space of 
the equivariant functions $ L_2(M; {\cal H}^\chi)^G $.

The projection operator 
$ P^\chi : L_2(M; {\cal H}^\chi) \to L_2(M; {\cal H}^\chi)^G $,
is now given by
\begin{equation}
	(P^\chi \psi)(x)
	= \int_G \rd g \, \rho^\chi(g) \psi(g^{-1}x).
	\label{P}
\end{equation}
{}From (\ref{red time}), (\ref{U(t)}), and (\ref{P})
the reduced time-evolution operator is given by
\begin{equation}
	( U^\chi (t) \psi ) (x) 
	= 
	\int_G \rd g \int_M \rd y \, 
	\rho^\chi(g) K(g^{-1}x,y;t) \psi(y)
\end{equation}
and thus the corresponding {\it reduced propagator} is
$ K^\chi : M \times M \times \R_{>0} \to \mbox{End} \, {\cal H}^\chi $
is defined by
\begin{equation}
	K^\chi(x,y;t)
	:= 
	\int_G \rd g \, \rho^\chi(g) K(g^{-1}x,y;t).
	\label{red pro}
\end{equation}
Our aim is to express the reduced propagator in terms of path integrals.

\section{Stratification geometry}
To write down a concrete form of the path integral
we need to equip the base space $ M $ with Riemannian structure.
Namely, now we assume that $ M $ is a differential manifold equipped with
a Riemannian  metric $ g_M $.
and that 
the Lie group $ G $ acts on $ M $ with preserving the metric $ g_M $.
Then the volume form induced from the metric defines an invariant measure
$ \rd x $ of $ M $.
We do {\it not} assume that the action of $ G $ on $ M $ is free.
Therefore $ p : M \to M/G $ is not necessarily a principal bundle.

{}For each point $ x \in M $,
$ G_x := \{ g \in G \, ; \, g x = x \} $ is called the 
{\it isotropy group} of $ x $
and
$ {\cal O}_x := \{ gx \, | \, g \in G \} $ is the 
{\it orbit} through $ x $.
It is easy to see $ {\cal O}_x \cong G/G_x $.
Note that the dimensions of the orbit $ {\cal O}_x $ can change suddenly
when the point $ x \in M $ is moved.
The subspace of the tangent space $ T_x M $,
$ V_x := T_x {\cal O}_x $, is called the 
{\it vertical subspace}
and its orthogonal complement
$ H_x := (V_x)^\perp $ is called the 
{\it horizontal subspace}.
$ P_V : T_x M \to V_x $ is the {\it vertical projection}
while
$ P_H : T_x M \to H_x $ is the {\it horizontal projection}.
A curve in $ M $ whose tangent vector always lies in the horizontal subspace
is called a {\it horizontal curve}.
Although these terms have been introduced
in the theory of principal fiber bundle,
we use them for a more general manifold that admits group action.

Let $ \Lie $ denote the Lie algebra of the group $ G $.
{}For each $ x \in M $,
$ \Lie_x $ is the Lie subalgebra of the isotropy group $ G_x $.
The group action $ G \times M \to M $ induces 
infinitesimal transformations $ \Lie \times M \to TM $ by differentiation.
The induced linear map
$ \theta_x : \Lie \to T_x M $ has
$ \mbox{Ker} \, \theta_x = \Lie_x $ and $ \mbox{Im} \, \theta_x = V_x $.
Then it defines an isomorphism
$ \widetilde{\theta}_x : \Lie/\Lie_x \isoarrow \, V_x $.
Now we define the {\it stratified connection form} $ \omega $ by
\begin{equation}
	\omega_x := (\widetilde{\theta}_x)^{-1} \circ P_V 
	: 
	T_x M \to \Lie/\Lie_x.
\end{equation}
Actually $ \omega $ is not smooth over the whole $ M $
but it is smooth on each stratification of $ M $.
\section{Reduction of path integral}
The Riemannian structure $ (M, g_M) $ 
defines the Laplacian $ \Delta_M $.
Suppose that $ V : M \to \R $ is a potential function such that
$ V(gx) = V(x) $
for all $ x \in M $, $ g \in G $.
Then
the Hamiltonian $ H = \dfrac12 \Delta_M + V(x) $,
which acts on $ L_2(M) $,
commutes with the action of $ G $, which is defined in (\ref{T}).
Let us accept that the path integral in $ M $ is formally given by
\begin{equation}
	K(x', x; t) 
	=
	\int_x^{x'} {\cal D}x 
	\exp
	\left[
		i \int_0^t \rd s 
		\left( 
			\frac{1}{2} ||\dot{x}(s)||^2 - V(x(s)) 
		\right)
	\right].
	\label{path int in M}
\end{equation}
Now we repeat our question;
what is the path-integral expression for the reduced propagator
(\ref{red pro}) on $ Q = M/G $?
The answer is our main result and is given below.
\vspace{3mm}
\\
{\bf Theorem 4.1.}
{\it The reduced path integral on $ Q = M/G $ is }
\begin{eqnarray}
	K^\chi(x',x;t) 
& = &
	\int_{q}^{q'} {\cal D}q \, 
	\rho^\chi ( \gamma )
	\nonumber 
	\\ && \quad
	\times \rho_*^\chi 
		\left(
			{\cal P} \exp
			\left[
				- \frac{i}{2} \int_0^t \rd s \,
				{\mit{\Lambda}} ( \widetilde{q}(s) )
			\right]
		\right)
	\nonumber 
	\\ && \quad
	\times \exp 
	\left[ 
		i \int_0^t \rd s 
		\left( \frac12 || \dot{q}(s) ||^2 - V(q(s)) \right)
	\right].
	\label{main}
\end{eqnarray}
\vspace{1mm}

{}To read the above equation we need explanation of the symbols.
The canonical projection map
$ p : M \to Q = M/G $
induces the metric $ g_Q $ of $ Q $ 
by asserting that the map $ p $ is a stratified Riemannian submersion.
For $ x, x' \in M $
we put $ q = p(x) $ and $ q'= p(x') $.
The map $ q : [0,t] \to Q $ is a curve connecting 
$ q = q(0) $ and $ q'= q(t) $.
The map $ \widetilde{q} : [0,t] \to M $ 
is a horizontal curve such that
$ \widetilde{q}(0) = x $ 
and 
$ p(\widetilde{q}(s)) = q(s) $ for $ s \in [0,t] $.
The element $ \gamma \in G $ is a holonomy defined by 
$ x' = \gamma \cdot \widetilde{q}(t) $.

To describe the symbol $ {\mit{\Lambda}} $,
which is called the {\it rotational energy operator},
we need more explanation.
The metric 
$ g_M : TM \otimes TM \to \R $ 
defines an isomorphism 
$ \widehat{g}_M : TM \to T^* M $.
Then its inverse map
$ \widehat{g}_M^{-1} : T^* M \to TM $
defines a symmetric tensor field
$ g_M^{-1} : M \to TM \otimes TM $.
Thus combining it with the stratified connection
$ \omega_x : T_x M \to \Lie / \Lie_x $ we define
the rotational energy operator by
\begin{equation}
	{\mit\Lambda} (x) := 
	- ( \omega_x \otimes \omega_x ) \circ g_M^{-1}(x)
	\in (\Lie / \Lie_x) \otimes (\Lie / \Lie_x).
\end{equation}
The unitary representation $ \rho^\chi $ of the group $ G $ 
over $ {\cal H}^\chi $
induces
the representation $ \rho^\chi_* $ 
of the universal envelop algebra $ {\cal U} (\Lie) $.
Then we have
$ \rho^\chi_* ( {\mit\Lambda} (x)) \in \mbox{End} \, {\cal H}^\chi $.
Moreover,
\begin{equation}
	\lambda(\tau) 
	=
	\rho_*^\chi 
	\left(
		{\cal P} \exp
			\left[
				- \frac{i}{2} \int_0^\tau \rd s \,
				{\mit{\Lambda}} ( \widetilde{q}(s) )
			\right]
	\right)
	\in \mbox{End} \, {\cal H}^\chi
\end{equation}
is defined as a solution of the differential equation
\begin{equation}
	\frac{d}{d \tau} \lambda(\tau) 
	=
	- \frac{i}{2} \,
	\rho_*^\chi 
	\left(
		{\mit{\Lambda}} ( \widetilde{q}( \tau ) )
	\right)
	\lambda(\tau),
	\qquad
	\lambda(0) = I \in \mbox{End} \, {\cal H}^\chi.
\end{equation}

Now we can read off the physical meaning of 
the reduced path integral (\ref{main}).
The path integral is expressed as a product of three factors:
\begin{enumerate}
\renewcommand{\labelenumi}{(\roman{enumi})}%
\item	the rotational energy amplitude
	$
		\exp
		[
			- \frac{i}{2} \int_0^t \rd s \,
			{\mit{\Lambda}} ( \widetilde{q}(s) )
		]
	$,
	which represents motion of the particle along the vertical directions
	of $ p : M \to M/G $;
\item	the vibrational energy amplitude
	$
		\exp 
		[ 
			i \int_0^t \rd s 
			( 
				\frac12 || \dot{q}(s) ||^2 - V(q(s)) 
			)
		]
	$,
	which represents motion of the particle along the horizontal directions;
\item	the holonomy factor $ \gamma $,
	which is caused by non-integrability of the horizontal distributions.
\end{enumerate}

Here we give the outline of the proof of the main theorem 4.1.
For the detail see the reference \cite{T}.
Essentially, it is only a matter of calculation;
from the path integral on $ M $ (\ref{path int in M})
\begin{equation}
	K(x', x; t)
	= 
	\int_x^{x'} {\cal D}x \, \re^{i I[x]},
	\qquad
	I[x] = 
	\int_0^t \rd s 
	\left( 
		\frac{1}{2} ||\dot{x}(s)||^2 - V(x(s)) 
	\right)
\end{equation}
with the reduction procedure (\ref{red pro}) we get
\begin{eqnarray}
	K^\chi(x',x;t) 
& := &
	\int_G \rd h \, \rho^\chi(h) K(h^{-1}x',x;t) 
	\nonumber \\
& = &
	\int_G \rd h \, \rho^\chi(h) 
	\int_x^{h^{-1}x'} {\cal D}x \, \re^{i I[x]} 
	\nonumber \\ 
& = &
	\int_G \rd h \, \rho^\chi(h) 
	\int_q^{q'} {\cal D}q
	\int_e^{h^{-1} \gamma} {\cal D}g
	\, \re^{i I[g \tilde{q} \, ]} 
	\nonumber \\ 
& = &
	\int_q^{q'} {\cal D}q
	\int_G \rd h 
	\, \rho^\chi(h) 
	\int_e^{h^{-1} \gamma} {\cal D}g
	\, \re^{i I[g \tilde{q} \, ]} 
	\nonumber \\ 
& = &
	\int_q^{q'} {\cal D}q
	\int_G \rd h 
	\, \rho^\chi(\gamma h) 
	\int_e^{h^{-1}} {\cal D}g
	\, \re^{i I[g \tilde{q} \, ]} 
	\nonumber \\ 
& = &
	\int_q^{q'} {\cal D}q
	\, \rho^\chi( \gamma )
	\int_G \rd h 
	\, \rho^\chi(h) 
	\int_e^{h^{-1}} {\cal D}g
	\, \re^{i \int \rd s  \frac12 ||\dot{g}||^2 } 
	\re^{i \int \rd s \{ \frac12 ||\dot{q}||^2 - V(q) \} } 
	\nonumber \\
& = &
	\int_{q}^{q'} {\cal D}q \, 
	\rho^\chi ( \gamma )
	\, 
	\rho_*^\chi 
		\left(
			{\cal P} \exp
			\left[
				- \frac{i}{2} \int_0^t \rd s \,
				{\mit{\Lambda}} ( \widetilde{q}(s) )
			\right]
		\right)
	\re^{i \int \rd s \{ \frac12 ||\dot{q}||^2 - V(q) \} }.
	\qquad
\end{eqnarray}

\section{Example}
Finally, we show an example of application of our formulation.
Let us begin with the plane $ M = \R^2 $,
which has the standard metric
$ g_M = \rd x^2 + \rd y^2 = \rd r^2 + r^2 \rd \theta^2 $.
It admits the symmetry action of 
$ G = SO(2) $.
The quotient space is a half line
$ Q = \R^2 / SO(2) = \Rplus $.
The invariant potential is a function $ V(r) $ only of $ r $.

The group action
\begin{equation}
	SO(2) \times \R^2 \to \R^2;
	\;
	\left(
		\begin{array}{rr}
		\cos \phi & - \sin \phi \\
		\sin \phi &   \cos \phi 
		\end{array}
	\right)
	\left(
		\begin{array}{c}
		x \\ y
		\end{array}
	\right)
\end{equation}
induces the action of the Lie algebra
\begin{equation}
	\LieSO(2) \times \R^2 \to T\R^2;
	\;
	\left(
		\begin{array}{rr}
		0    & - \phi \\
		\phi &   0      
		\end{array}
	\right)
	\left(
		\begin{array}{c}
		x \\ y
		\end{array}
	\right),
\end{equation}
which defines the vertical distribution
\begin{equation}
	\theta : \LieSO(2) \times \R^2 \to T\R^2;
	\;
	\left(
	\left(
		\begin{array}{rr}
		0    & - \phi \\
		\phi &   0      
		\end{array}
	\right),
	\,
	\left(
		\begin{array}{c}
		x \\ y
		\end{array}
	\right)
	\right)
	\mapsto
	\phi \frac{\partial}{\partial \theta}.
\end{equation}
Then the stratified connection becomes
\begin{equation}
	\omega =
	\left(
		\begin{array}{rr}
		0 & -1 \\
		1 &  0   
		\end{array}
	\right)
	\rd \theta.
\end{equation}
In the cotangent space the metric is given as
\begin{equation}
	(g_M)^{-1} = 
	\dfrac{\partial}{\partial r} \otimes
	\dfrac{\partial}{\partial r}
	+
	\dfrac{1}{r^2} 
	\dfrac{\partial}{\partial \theta} \otimes
	\dfrac{\partial}{\partial \theta}.
\end{equation}
The rotational energy operator is
\begin{equation}
	{\mit{\Lambda}}
	= - ( \omega \otimes \omega ) \circ (g_M)^{-1}
	=
	- \dfrac{1}{r^2}
	\left(
		\begin{array}{rr}
		0 & -1 \\
		1 &  0   
		\end{array}
	\right)
	\otimes
	\left(
		\begin{array}{rr}
		0 & -1 \\
		1 &  0   
		\end{array}
	\right).
\end{equation}
The irreducible unitary representations of $ SO(2) $ are
labeled by an integer $ n \in \Z $ and defined by
\begin{equation}
	\rho_n : SO(2) \to U(1);
	\;
	\left(
		\begin{array}{rr}
		\cos \phi & - \sin \phi \\
		\sin \phi &   \cos \phi 
		\end{array}
	\right)
	\mapsto
	\re^{i n \phi}.
\end{equation}
The differential representation of the Lie algebra of $ SO(2) $
is 
\begin{equation}
	(\rho_n)_* : \LieSO (2) \to \LieU (1);
	\;
	\left(
		\begin{array}{rr}
		0 & - \phi \\
		\phi &  0 
		\end{array}
	\right)
	\mapsto
	i n \phi.
\end{equation}
The rotational energy operator is then represented as
\begin{equation}
	(\rho_n)_* ( {\mit{\Lambda}} )
	=
	- \dfrac{1}{r^2} ( in )^2
	=
	\dfrac{n^2}{r^2}.
\end{equation}
Finally the reduced path integral is given by
\begin{equation}
	K_n (r', \theta', r, \theta ;t) 
	= 
	\int_{r}^{r'} {\cal D} r 
	\, \re^{ in ( \theta' - \theta) }
	\exp \!
	\left[
		i \! \int_0^t \! \rd s
		\bigg\{
			- \frac{n^2}{2 r^2} 
			+ \frac12 \dot{r}^2 
			- V(r)
		\bigg\}
	\right].
\end{equation}
So the effective potential for the radius coordinate $ r $ is given by
\begin{equation}
	V_{\mathrm{eff}}(r) = V(r) + \frac{n^2}{2 r^2},
\end{equation}
where the second term represents the centrifugal force.

\section*{Acknowledgments}
Thanks are due organizers for the stimulating conference at the beautiful site,
Sts. Constantine and Elena in Bulgaria.
I wish to thank
I. M. Mladenov 
especially
for his effort to keep touch with me for a long time from a great distance.
I have benefited also from discussions with 
P. Exner and C. Herald.
This work is supported by a Grant-in-Aid for Scientific Research 
({\#}12047216)
from
the Ministry of Education, Culture, Sports, Science and Technology of Japan.


\begin{thebibliography}{99}
\bibitem{Davis}
	Davis M,
	{\it Smooth G-manifolds as Collections of Fiber Bundles},
	Pacific J. Math. {\bf 77} (1978) 315--363.
\bibitem{Landsman-Linden}
	Landsman N P, Linden N,
	{\it The Geometry of Inequivalent Quantizations},
	Nucl. Phys. B {\bf 365} (1991) 121--160.
\bibitem{M}
	Montgomery R,
	{\it Isoholonomic Problems and Some Applications},
	Commun. Math. Phys. {\bf 128} (1990) 565--592.
\bibitem{TT}
	Tanimura S, Tsutsui I,
	{\it Induced Gauge Fields in the Path-Integral},
	Mod. Phys. Lett. A {\bf 10} (1995) 2607--2617;
	e-print archive: 
	hep-th/9508165.
\bibitem{TI}
	Tanimura S, Iwai T,
	{\it Reduction of Quantum Systems on Riemannian Manifolds
	with Symmetry and Application to Molecular Mechanics},
	J. Math. Phys. {\bf 41} (2000) 1814--1842;
	e-print archive: 
	math-ph/9907005.
\bibitem{T}
	Tanimura S,
	{\it Path Integrals on Riemannian Manifolds with Symmetry
	and Induced Gauge Structure},
	Int. J. Mod. Phys. A {\bf 16} (2001)  1443--1461;
	e-print archive: 
	hep-th/0006150.
\end{thebibliography}
\end{document}